\documentstyle[editedvolume,psfig]{crckapb}
\begin{opening}
\title{Physical Parameters in Relativistic Jets from Compact Symmetric Objects}
\author{M. Perucho and J.M. Mart\'i}
\institute{Dpt. Astronomia i Astrof\'{\i}sica, Universitat de Val\`encia\\
46100 Burjassot (Val\`encia), SPAIN}

\end{opening}  
\begin{document}

\maketitle

\begin{abstract}
\small{
{\it Compact symmetric objects} conform a class of sources characterized 
by high luminosity radio emission located symmetrically on both sides of the 
active galactic nucleus on linear scales of less than 1 kpc. Given their small 
size, the hot spots of the jets in CSOs provide a unique 
laboratory for the study of the physics of relativistic jets and their 
environment close to the central engine.\\
  We present a simple model for the hot spots in CSOs assuming synchrotron 
emission, minimum energy and ram pressure equilibrium with the external medium. 
Further comparison of our model with observational data allows us to constrain 
the physical parameters in the hot spots and the jets feeding them, and the 
density profile of the external medium.}
\end{abstract}

\section{Introduction}
  In the early eighties Phillips and Mutel (1982) first detected 
compact radio sources with double structure, known as {\it compact doubles}. 
Later on, {\it compact symmetric objects} (CSOs) were defined by Wilkinson et 
al. (1994) as core-detected sources with double sided emission and total linear 
size smaller than 1 kpc. They present a typical synchrotron spectrum at high 
radio frequencies, with a peak due to absorption, around the GHz, i.e., they are 
also included within a class of objects known as {\it gigahertz peaked spectrum 
sources}.

  The main controversy about the nature of CSOs has been if they are old, 
confined sources (van Breugel et al. 1984; Carvalho 1998) or the young 
precursors of large Faranoff-Riley II (Owsianik et al. 1998; Taylor et al. 
2000). The study of this class of compact sources is thus very interesting as a 
family of objects on their own, and they provide a scenario to probe jet 
properties close to the active nucleus.

  In this contribution, we present results for the physical conditions in the 
jets and hot spots in a sample of CSOs and summarize the main conclusions of a 
simple evolution model, trying to discern between the proposed scenarios by
comparing the model with observational data. All the quantities have been 
calculated assuming synchrotron emission, minimum energy and ram pressure 
equilibrium with the external medium. Calculations are performed assuming 
the Universe is described by the standard Friedmann-Robertson-Walker model with
Hubble constant $H_{0} = 70$ km s$^{-1}$ Mpc$^{-1}$ and deceleration 
parameter $q_{0} = 0.5$. 

  Sources have been selected from the GPS samples of Stanghellini et al. (1998), 
Snellen et al. (1998), and Peck et al. (2000). We have chosen those sources with double
morphology and already classified in the literature as CSOs and also those whose
components can be safely interpreted as hot spots even though the central core
has not been yet identified. Moreover, sources possibly affected by orientation
effects (beaming, spectral distortion), like quasars and core-jet sources, have
not been considered. The resulting sample is formed by 20 sources which are
listed, along with the data relevant for our study, in Perucho \& Mart\'{\i} 
(2001).

\section{Physical parameters of hot spots and jets}   
  Panels (a) and (b) display hot spot radius ($r_{\rm hs}$) and hot spot 
luminosity ($L_{\rm hs}$), respectively, versus source linear size ($LS$). These 
quantities are directly obtained from the corresponding  
angular sizes and the formulas for cosmological distance. A selfsimilar 
evolution for the hot spot ($r_{\rm hs} \propto LS$) is clearly observed. 
The hot spot luminosity seems to be independent of the source linear size, with
only a weak tendency to grow with $LS$. 

  We have also estimated the internal energy and pressure in the hot spots  
assuming that the magnetic field strength and the particle energy distribution 
arrange in the most efficient way to produce the estimated (synchrotron)
luminosity. These minimum energy conditions are reached near the equipartition 
conditions in which magnetic field and particle energy have the same values.
Finally, the number density of relativistic particles within the
hot spots is estimated asuming that there is no thermal (leptonic nor barionic) 
component. Panels (c) and (d) show the distribution of pressures and 
number densities of relativistic particles in the hot spots, $p_{\rm hs}$ and 
$n_{\rm hs}$, respectively, as a function of $LS$. We refer the interested 
reader to the Appendix in Perucho \& Mart\'{\i} (2001) for details on our 
model.

\begin{figure}[h]
  \centerline{%
  \psfig{bbllx=35pt,bblly=370pt,bburx=575pt,bbury=739pt,%
       file=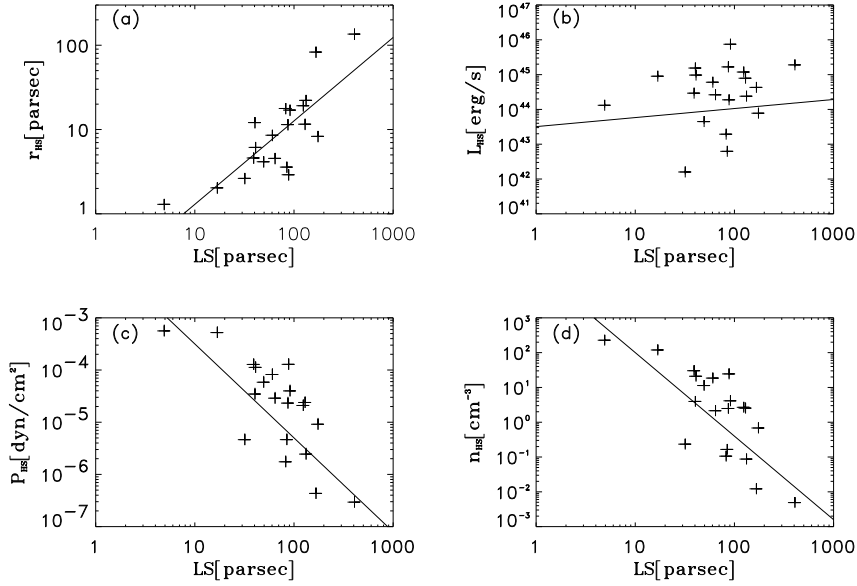,width=\textwidth,angle=0,clip=}%
  }
 \tiny{
\caption{Mean radius (panel a), radio luminosity (b), pressure (c) and electron 
density (d) of hot spots versus linear size. Solid lines correspond to the best
linear log-log fits. Results are $\log(r_{\rm hs}) \propto (1.0\pm0.3)\log(LS)$ 
($r=0.8$), $\log(L_{\rm hs}) \propto (0.4\pm0.5)\log(LS)$ ($r=0.2$), 
$\log(P_{\rm hs}) \propto (-1.7\pm0.4)\log(LS)$ ($r=-0.75$), and  
$\log(n_{\rm hs}) \propto (-2.3\pm1.3)\log(LS)$ ($r=-0.75$). $r$ stands for the
regression coeficient of the corresponding fits. Error bars correspond to 
15\% errors in angular size and measured radio flux and are just indicative.}}
\label{figure1}
\end{figure}%

  We can use the values of the pressure and particle number density within 
the hot spots to estimate the fluxes of momentum (thrust), energy and particles 
in the relativistic jets feeding them. Ram pressure equilibrium between the
jet and hot spot leads to $F_{\rm j} = P_{\rm hs}A_{\rm hs}$, for the jet
thrust $F_{\rm j}$, where $A_{\rm hs}$ stands for the hot spot cross section
($\simeq \pi r_{\rm hs}^2$). Taking mean values for $P_{\rm hs}$ and 
$r_{\rm hs}$ from our sample we get $F_{\rm j} \simeq 8.3\,10^{34}$ dyn. In a
similar way, the particle flux in the jet, $R_{\rm hs}$, can be estimated from
the total number of particles in the hot spot, $n_{\rm hs}V_{\rm hs}$ 
($V_{\rm hs}$ is the hot spot volume, $\simeq 4 \pi r_{\rm hs}^3/3$) and the
source lifetime, $\simeq v_{\rm hs}/LS$, where $v_{\rm hs}$ is the hot spot
advance speed (assumed constant and $\simeq 0.2c$; see below). According to 
this, $R_{\rm j} = n_{\rm hs}V_{\rm hs}v_{\rm hs}/LS \simeq 
1.3\,10^{49}$ e$^{+/-}$ s$^{-1}$. Finally, a lower bound for the jet power, 
$Q_{\rm j}$, can be estimated considering that, in a relativistic jet, 
$Q_{\rm j} = (F_{\rm j}/v_{\rm j}) c^2$. Hence, for given $F_{\rm j}$ and 
taking $v_{\rm j} = c$, we have $Q_{\rm j, min} = F_{\rm j}c =
P_{\rm hs} A_{\rm hs} c = 2.5\,10^{45}$ erg s$^{-1}$. It is interesting to note
that the lower bound for the jet power is consistent with the 
one obtained by Rawlings \& Saunders (1991) for FRII radio galaxies ($10^{44}$ 
erg s$^{-1}$) supporting the idea of CSOs being the early phases of FRIIs.
Also, the flux of particles inferred in the jet is consistent with 
accretion rates of barionic plasma of the order $0.35\,M_\odot$ y$^{-1}$, 
implying a highly efficient conversion of accretion energy at the Eddington 
limit into ejection (or a leptonic composition of jets).

\section{An evolution model for CSOs}
  The results presented in Sect.~2 admit an evolutive interpretation assuming 
the linear size of the source ($LS$) is related with time by means of the
hot spot advance speed ($v_{\rm hs}$). According to this, we have developed
a model (Perucho \& Mart\'{\i} 2001) for the evolution of CSOs within the first 
kpc assuming that the advance speed is determined by ram-pressure equilibrium 
between the relativistic jet and the external medium and consistent with the 
self-similar character of the source growth. The linear fits shown in Fig.~1
together with the assumption of an almost constant advance speed (Owsianik et 
al. 1998, Taylor et al. 2000), allow us to determine the values of the 
parameters in our model. Results show that, in order to mantain increasing 
luminosity and constant advance speed, it is required to invest increasing 
power with time which can lead to severe energy problems. On the other hand,
fixing a constant jet power supply, keeping constant advance speed requires 
decreasing luminosity, and viceversa: when luminosity is kept increasing with 
linear size, external density has a less steep gradient, turning out in a less 
steeply decreasing internal pressure, and a decreasing advance speed; on the 
other hand, keeping speed constant, external density has to decrease faster, 
causing a fast expansion of the hot spot and a decrease in luminosity.

  Finally, our results allow us to estimate the ages and speeds of the sources 
at one kpc from the central engine. Considering a (constant) advance 
speed of $0.2c$, the age is $1.6\,10^4$ y. If we allow for a decrease in 
the advance speed (as suggested by one of our models), the age of a 
source which starts to decelerate at 10 pc (100 pc) will be of $10^5$ y
($3.5\,10^4$ y) at 1 kpc. These results support the {\it young source scenario}
for CSOs. Moreover, the speeds of the hot spots at the end of the first kpc
($0.02 - 0.06c$) are consistent with the terminal speeds inferred for FRIIs.

  Much more conclusive answers could be given if there was a larger sample of 
CSOs with i) measured advance speeds, and ii) measured luminosities.

\end{document}